\documentclass{WileyMSP-template}

\usepackage{amsmath} 
\usepackage{hyperref}
\usepackage{dcolumn}
\usepackage{graphicx,color}
\usepackage[utf8]{inputenc}
\usepackage{amsmath,amssymb,bm,amsfonts,dsfont,mathrsfs,amsthm}
\usepackage{braket}
\usepackage{txfonts,comment}
\usepackage{soul}
\usepackage{textcomp}
\usepackage{etoolbox}
\usepackage{multirow}
\usepackage{bm}
\usepackage{bbold}
\usepackage{yhmath}
\usepackage[english]{babel}
\usepackage[utf8]{inputenc}
\usepackage{xcolor}

\renewcommand{\k}[1]{\vert #1\rangle}
\renewcommand{\b}[1]{\langle #1\vert}
\newcommand{\up}{\vert\uparrow\rangle}
\newcommand{\down}{\vert\downarrow\rangle}

\renewcommand{\mod}[1]{\vert #1 \vert}
\newcommand{\betau}{\vert\beta_\uparrow\rangle}
\newcommand{\betad}{\vert\beta_\downarrow\rangle}

\begin{document}


\title{Phase information transfer by post-selection in Spin--Mechanical assisted magnetometry}

\maketitle

\author{Ra\'ul Coto*,}
\author{Hugo Molinares,}
\author{Vitalie Eremeev}

\begin{affiliations}

Ra\'ul Coto\\
Department of Chemistry and Physics, Nova Southeastern University, Fort Lauderdale-Davie, Florida 33328-2004, USA\\
Email Address: rcotocab@nova.edu

\medskip
Hugo Molinares\\
Departamento de Ciencias F\'{\i}sicas, Universidad de La Frontera, Casilla 54-D, Temuco, Chile

\medskip
Vitalie Eremeev\\
Centro Multidisciplinario de F\'isica, Vicerrector\'ia de Investigaci\'on, Universidad Mayor, 8580745 Santiago, Chile

\end{affiliations}

\justifying

\begin{abstract}
Quantum information transfer between light-matter-type systems is poised to enable important applications, while also serving as a testbed for theoretical investigation. Such systems can be realized with spins coupled to a mechanical oscillator, a platform that has been extensively studied both theoretically and experimentally. Early demonstrations of quantum information transfer have relied mostly on coherent control. However, measurement-induced backaction has emerged as a strong alternative for quantum control. In this work we use post-selection on the spin system as a selective backaction to refocus spin's phase information onto the mechanical oscillator. We identify physical resources and operating regimes that govern conditional phase transfer, including oscillator quantum coherence, the number of spins, the mechanical initial state, coupling strength, oscillator amplitude, and relaxation. We benchmark different scenarios using the variance as the figure of merit, estimated via two complementary approaches: a semiclassical variance estimator and a Pegg-Barnett quantum estimator. The Cram\'er-Rao bound is also computed for comparison. The analysis provides a framework for understanding phase transfer in high-dimensional hybrid quantum systems and for measurement-induced backaction used in quantum magnetometry.

\end{abstract}

\justifying

\section{Introduction}\label{introduction}

Quantum information transfer is a central topic in quantum science, with applications in quantum communication, distributed quantum computation, quantum control, and quantum sensing. Early protocols focused primarily on the coherent transfer of quantum states between distant nodes~\cite{Lemonde_2018,Kandel_2019}, while other studies investigated the transfer of correlations, coherences, and populations between different physical subsystems~\cite{Nakajima_2018,Peng_2025,Coto_2017B,Gonzalez_2022B}. These developments have established coherent control as a powerful mechanism for manipulating quantum information. However, measurement-induced backaction provides an alternative route in which quantum information can be redirected, conditioned, or amplified through selective measurements~\cite{Blok,Montenegro2}.

In this context, post-selection has attracted significant attention for being a paradigm in which measurement outcomes are selectively retained~\cite{Aharonov}. Post-selection has been shown to be advantageous in numerous experiments including optical beam deflection \cite{Dixon,Starling}, polarization measurements \cite{Pryde}, interferometric phase estimation \cite{Brunner} among others~\cite{Kocsis,Feizpour,Zhang,Alves_exp}. In parallel, several theoretical studies have investigated the metrological advantages for parameter estimation \cite{Dressel,Torres,Coto,Magana,Arvidsson-Shukur, Coto_2021}.


Spin--mechanical systems provide a natural platform for studying quantum information transfer via post-selection. They combine a well-controlled two-level system with a mechanical oscillator possessing a high-dimensional Hilbert space. In these systems, magnetic-gradient coupling, strain coupling, or other dispersive interactions can correlate spin states with mechanical displacement~\cite{Rabl_2008,Arcizet_2011,Kolkowitz_2012,Bennett_2012,Lee_2016}.

In this work we investigate phase-information transfer via post-selection in a hybrid spin--mechanical quantum magnetometry protocol. We consider a spin system whose phase encodes a magnetic field and transfer phase-information to a mechanical oscillator (MO). We identify quantum coherence~\cite{Baumgratz_2014} and the number of spins participating in the protocol as resources that benefit this process. In addition, we compare different preparations of the MO initial state, including coherent states, cat states, and thermal states. These states provide different balances between phase-space localization and nonclassical interference. 

We further study the effects of spin relaxation, spin--mechanical coupling strength, and oscillator amplitude. To quantify the performance of the process, we evaluate the uncertainty carried by the phase using the variance as the figure of merit. We obtain this variance in two complementary approaches. First, we use a semiclassical error-propagation analysis, commonly used in spin-magnetometry experiments \cite{Maze,Aiello}. Second, we use the quantum phase-operator formalism introduced by Pegg and Barnett \cite{Pegg}, which provides a fully quantum description of phase measurements.

The paper is organized as follows. In Sec.~\ref{Sec_spin_mechanical model} we describe the system, its evolution and the main features of the protocol. In Sec.~\ref{Sec_Variance} we introduce the variance's estimators used to benchmark the phase transfer. In Sec.~\ref{Sec_results} we organize and discuss our results. In Sec.~\ref{Sec_conclusions} we provide the final remarks of this work.

\section{Spin--mechanical model}\label{Sec_spin_mechanical model}

We consider a hybrid quantum system consisting of a spin-qubit coupled to a mechanical oscillator (MO). In a frame rotating at the spin frequency, the dynamics of the coupled system can be modeled by a conditioned-displacement Hamiltonian (with $\hbar = 1$)
\begin{equation}\label{Hsys} 
\hat{H} = \frac{\gamma B_z}{2}\hat{\sigma}_z + \omega_m \hat{b}^\dag \hat{b} + \lambda_0 \hat{\sigma}_z (\hat{b}^\dag  +  \hat{b}),
\end{equation}
 where $\gamma$ is the gyromagnetic ratio, $B_z$ is a static magnetic field, $\omega_m$ is the frequency of the mechanical mode,  $\lambda_0$ is the spin-mechanics coupling strength, $\hat{b}$ ($\hat{b}^\dag$) stands for the annihilation (creation) bosonic operator and $\hat{\sigma}_z$ is the Pauli operator. The unitary evolution operator $\hat{U}(t)=e^{-i\hat{H}t}$, can be conveniently factorized by applying a transformation defined through $\hat{H}^\prime=e^{\hat{S}}\hat{H}e^{-\hat{S}}$, with $\hat{S}=\lambda_0/\omega_m \hat{\sigma_z}(\hat{b}^\dagger - \hat{b})$. Within this framework, the time-evolution operator can be expressed as \cite{Bose_97,Montenegro1} 
 \begin{equation}\label{Uop}
     \hat{U}(t)=\exp{[-i\gamma B_z\hat{\sigma}_z t/2]}\exp{[-\lambda\hat{\sigma}_z(\xi\hat{b}^\dagger - \xi^\ast\hat{b})]}\exp{[-i\omega_m\hat{b}^\dagger\hat{b}t]},
 \end{equation} 
 where $\xi=1-e^{-i\omega_mt}$, and $\lambda=\lambda_0/\omega_m$ denotes the dimensionless spin-mechanics coupling strength. 
 
 Consider the system initially prepared in a product state where the spin is in a balanced superposition of the eigenstates of the operator $\hat{\sigma}_z$ and the MO is in a coherent state $\ket{\beta}$, expressed as $\k{\Psi_i}=\big[\cos(\theta_i/2)\up + \sin(\theta_i/2)\down\big]\k{\beta}$. We use the coherent state for analytical calculations and resort to numerical calculations for other initial states of the MO. The bipartite system evolves under the unitary evolution operator introduced above during a sensing time $\tau$. Typically, sensing protocols rely on the accumulation of a phase proportional to the magnetic field during the time evolution of a spin probe~\cite{Maze,Aiello,Degen,Zaiser}. In addition, correlations between the subsystems develop, leading in general to spin-mechanics entanglement. The time $\tau$ governs how much of the spin state is imprinted onto the MO, providing direct control over the transfer process. The joint spin-mechanics state (up to a global phase) after the evolution is then given by
\begin{equation}\label{composite_state}
\k{\Psi}=\cos(\theta_i/2)e^{-i(\varphi+2\lambda\beta\sin(\omega_m\tau))}\up\betau+ \sin(\theta_i/2)\down\betad,
\end{equation} 
where  $\betau =\k{\beta e^{-i\omega_m\tau} - \lambda\xi}$ and $\betad = \k{\beta e^{-i\omega_m\tau} + \lambda\xi}$ are the displaced coherent states, and $\varphi=\gamma B_z\tau$. Note that by measuring the resulting phase $\varphi$ one can estimate the amplitude of the magnetic field $B_z$.

It is known that the maximum amount of information about the parameter $\varphi$ that can be extracted from measurements performed on a mixed state $\rho(\varphi)$ is quantified by the quantum Fisher information (QFI) as follows \cite{Paris2009}:
\begin{equation}\label{qfi_g}
\mathcal{F}(\varphi)=2\sum_{n,m}\frac{\left|\left\langle \Phi_m \left| \partial_\varphi \rho(\varphi) \right| \Phi_n \right\rangle\right|^2}{\epsilon_n+\epsilon_m},
\qquad
\epsilon_n+\epsilon_m \neq 0.
\end{equation}
Here, the spectral decomposition of the density operator is given by $\rho(\varphi)=\sum_n\epsilon_n\ket{\Phi_n}\bra{\Phi_n}$ where $\epsilon_n$ and $\ket{\Phi_n}$ denote the $n$th eigenvalue and eigenvector of $\rho(\varphi)$, respectively. Furthermore, $\partial_\varphi \equiv \partial/\partial\varphi$ represents the derivative with respect to the phase parameter $\varphi$. Equation (\ref{qfi_g}) provides the general expression for the QFI of an arbitrary quantum state. In the particular case of a pure state, $\rho(\varphi)=\ket{\Psi(\varphi)}\bra{\Psi(\varphi)}$, the QFI reduces to \cite{Fisher,Helstrom,Alves_exp,Alves,Arvidsson-Shukur}
\begin{equation}\label{QFI}
\mathcal{F}(\varphi)=4\left(\langle \partial_\varphi\Psi|\partial_\varphi\Psi\rangle
-\left|\langle \Psi|\partial_\varphi\Psi\rangle\right|^2\right).
\end{equation}

 Substituting Eq.(\ref{composite_state}) into Eq.(\ref{QFI}) yields $\mathcal{F}(\varphi)=\sin^2(\theta_i)$. This result shows that the maximum amount of information that can be extracted about the phase $\varphi$ is bounded by $\sin^2(\theta_i)$, independently of the value of $\varphi$. The optimal preparation is therefore achieved for $\theta_i=\pi/2$, corresponding to the spin in a coherent superposition state.

We now consider a post-selection of the spin onto the state $\k{\psi_p}=\cos(\theta_p/2)\up +\sin(\theta_p/2)e^{-i\phi_p}\down $. Conditioned on a successful post-selection event, the MO is projected onto the state
\begin{equation}\label{mechanical_state}
\k{\phi_m}=\alpha_{\uparrow}\betau + \alpha_{\downarrow}\betad,
\end{equation}
where $\alpha_{\uparrow}=\cos(\theta_i/2)\cos(\theta_p/2)e^{-i(\varphi+2\lambda\beta\sin (\omega_m \tau) )}/\sqrt{p(\varphi)}$, $\alpha_{\downarrow}=\sin(\theta_i/2)\sin(\theta_p/2)e^{i\phi_p}/\sqrt{p(\varphi)}$, and the probability of a successful post-selection is given by
\begin{eqnarray}
p(\varphi)=\frac{1}{2}\left[ 1 + \cos(\theta_i)\cos(\theta_p) +  \sin(\theta_i)\sin(\theta_p)\cos\big(\varphi+\phi_p+4\beta\lambda\sin (\omega_m \tau)\big) e^{-2\lambda^2\mod{\xi}^2}\right].
\end{eqnarray}
Unless otherwise stated, we set $\theta_p=1.65$, $\phi_p=0$, $\theta_i=\pi/2$, $\lambda=0.1$ and $\gamma/\omega_m=1$ G$^{-1}$. In Appendix~\ref{appendix_fisher} we compare the QFI obtained in this post-selection scenario with the unpostselected case discussed above.

\section{Variance's estimators}\label{Sec_Variance}

In this section we introduce two ways to estimate the variance of the phase, which is directly related to the magnetic field.

\subsection{Semiclassical phase variance}

We first consider a widely used error propagation analysis to estimate the variance of the phase in terms of the $\hat{X}$ quadrature. It can be expressed as~\cite{Maze,Aiello,Ivanov}
\begin{equation}
\Delta\varphi_{SC}^2=\frac{\langle \Delta \hat{X}\rangle^2}{\left|\partial_\varphi  \langle \hat{X}\rangle \right|^2},
\end{equation}
where $\langle\Delta \hat{X}\rangle^2=\langle \hat{X}^2\rangle-\langle \hat{X}\rangle^2$, and the expectation values are calculated with respect to the post-selected state $\langle \phi_m|\hat{X}|\phi_m\rangle$, with the quadrature operator defined as $\hat{X}=(\hat{b}+\hat{b}^\dagger)/2$. In the Appendix~\ref{appendix_calculations} one can note that $\langle \hat{X}\rangle$ exhibits a sharp variation with respect to the phase $\varphi$, which could lead to a reduced variance $\Delta\varphi_{SC}^2$. This apparent enhancement in sensitivity is often concluded as one of the main motivations for considering strategies based on post-selection. We further elaborate on this idea in the Appendix.

\subsection{Quantum (Pegg-Barnett) phase variance}
An alternative approach to estimating the phase variance is based on the construction of a proper quantum phase operator. We adopt the Pegg–Barnett formalism \cite{Pegg}, in which the Hermitian phase operator is defined as $\hat{\varphi}=\sum_{j=0}^s\Theta_j\k{\Theta_j}\b{\Theta_j}$, where the phase states are given by $\k{\Theta_j}=\lim\limits_{s\rightarrow \infty} (1+s)^{-1/2}\sum_{n=0}^{s}e^{in\Theta_j}\k{n}$. These states form an orthonormal basis expressed in terms of the Fock states $\k{n}$. The discrete phase values are defined as 
$\Theta_j =\Theta_0 +2\pi j/(s+1)$, 
where $\Theta_0$ is an arbitrary reference phase and $j=0,\dots,s$. Within this framework, the phase properties of a quantum state can be consistently described in the limit $s\rightarrow\infty$ (this will be considered later). For a coherent state $\k{\alpha}$, with $\alpha = r e^{i\nu}$, the corresponding phase amplitude can be obtained:
\begin{equation}
\b{\Theta_j}\alpha\rangle =e^{-r^2/2}(s+1)^{-1/2}\sum_{n=0}^{s}\frac{r^n}{\sqrt{n!}}e^{i(\nu-\Theta_j)n}.
\end{equation}
We now proceed to evaluate the expectation value $\langle \hat{\varphi} \rangle=\b{\alpha}\hat{\varphi}\k{\alpha}$ and the corresponding variance $\Delta\hat{\varphi}_{PB}^2 =(\langle \hat{\varphi}^2 \rangle - \langle \hat{\varphi} \rangle^2)/\left|\partial_\varphi  \langle\hat{\varphi}\rangle \right|^2$. The expectation values of the powers of the phase operator are given by  $\langle \hat{\varphi}^k \rangle =\sum_{j}\Theta_j^k\mod{\b{\Theta_j}\alpha\rangle}^2$. By replacing the coherent state $\k\alpha$ with the post-selected mechanical state $\k\phi_m$ defined in Eq.~(\ref{mechanical_state}), we obtain
\begin{eqnarray}
\label{meanPB}
\langle \hat{\varphi}^k \rangle &=& \vert \alpha_\uparrow \vert^2 \sum_{j}^s\Theta_j^k\mod{\b{\Theta_j}\beta_\uparrow\rangle}^2+ \vert \alpha_\downarrow \vert^2 \sum_{j}^s\Theta_j^k\mod{\b{\Theta_j}\beta_\downarrow\rangle}^2 + \left(\alpha_\uparrow \alpha_{\downarrow}^\ast \sum_{j}^s\Theta_j^k \b{\beta_\downarrow}\Theta_j \rangle\b{\Theta_j}\beta_\uparrow\rangle + h.c.\right),
\end{eqnarray}
where $k=\{1,2\}$. In the limit $s \rightarrow \infty$, the discrete summation appearing in Eq.(\ref{meanPB}) can be replaced by an integral over the phase variable $\Theta = 2\pi(k- s/2)/(s+1)$, with the integration domain extending from $-\pi$ to $\pi$~\cite{Lynch}.

Foundational works have established the ultimate precision bounds achievable with quantum probes and measurements, typically formulated in terms of the Quantum Fisher Information (QFI) and the corresponding Cram\'er-Rao bound~\cite{Cramer,CRao}. Comprehensive reviews of the field can be found in Refs. \cite{Giovannetti2004,Giovannetti2011,Degen}. The above definitions of the variance do not explicitly account for the success probability associated with the post-selection process. Hence, comparing to the Cram\'er-Rao bound, we only consider the QFI corresponding to the post-selected state $\mathcal{F}_m(\varphi)$, see Appendix~\ref{appendix_fisher}. In this case, the bound reads $\Delta\varphi_{\mathrm{CR}}^{2}=1/\mathcal{F}_m(\varphi)$. 

\section{Results}\label{Sec_results}

\subsection{Attainable variance with quantum coherence witness}

\begin{figure}[ht]
\centering
\includegraphics[scale=0.6]{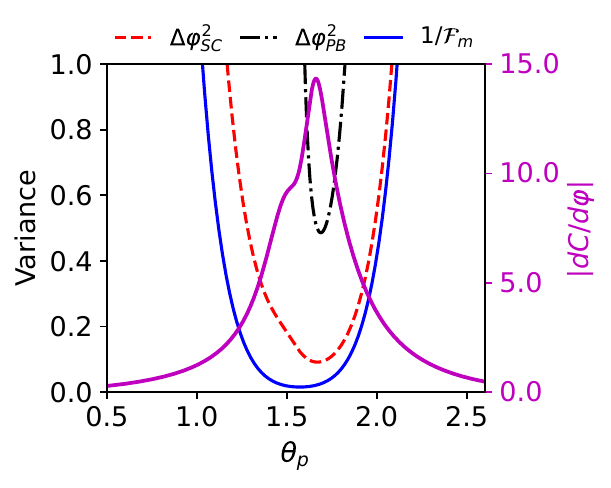}
\includegraphics[scale=0.6]{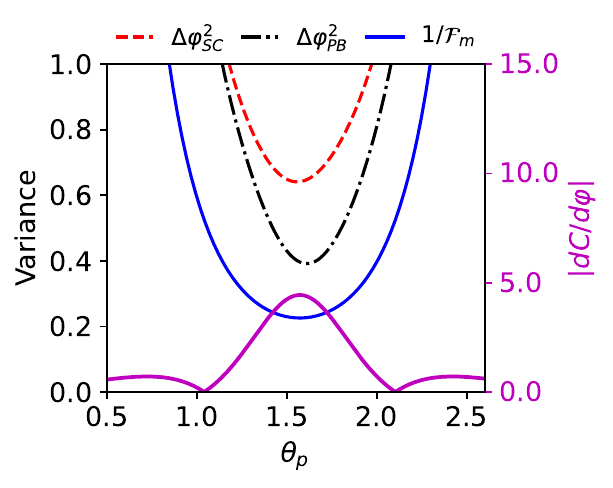}
\includegraphics[scale=0.6]{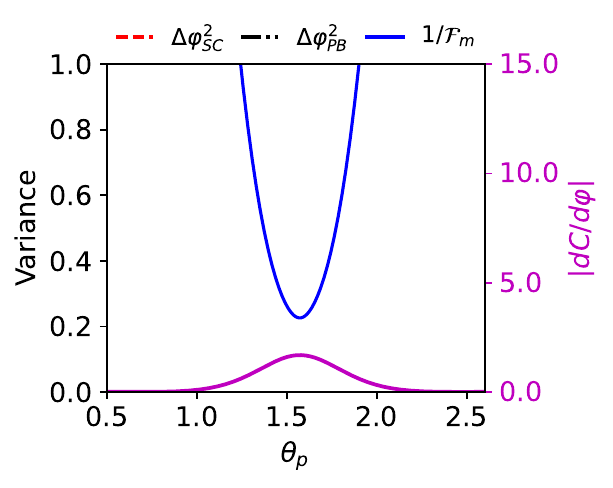}
\caption{Left panel: coherent-state $\ket{\beta}$ with amplitude $\beta=1$, center panel: even cat-state $\ket{\mathrm{cat}}\propto(\ket{\alpha}+\ket{-\alpha})$, with $\vert\alpha\vert=2$; and right panel: thermal-state with mean occupation $\bar{n}_{\mathrm{th}}=0.5$. The phase variance is estimated using two different approaches: a semiclassical procedure based on the error-propagation formula $(\Delta \varphi_{SC}^{2})$ and the Pegg-Barnett formalism for the phase operator $(\Delta \hat{\varphi}_{PB}^{2})$. $1/\mathcal{F}_m$ (CR) represents the lowest bound. Parameters used are $\lambda=0.1$, $\omega_m\tau=1$, $\theta_i=\pi/2$, $\gamma/\omega_m=1$ G$^{-1}$, and $B_z = 2.7$ G.}
\label{fig_variance}
\end{figure}

We devote this section to analyze the variance in different scenarios. We consider two estimators of the variance, as detailed in Sec.~\ref{Sec_Variance}, namely: semiclassical (SC) and Pegg-Barnett (PB). For comparison, we also consider the Cramer-Rao bound (CR). Throughout numerical simulations we observe consistency among the estimators. However, small differences arise depending on the initial state of the MO. Comparing these two estimators allows us to identify the regimes in which a semiclassical phase estimator is reliable and those in which the discrete, quantum nature of the oscillator phase becomes important. In Fig.~\ref{fig_variance} we observe that for the MO initially in a coherent state, the semiclassical approach provides a good estimation---we stress that CR is the lowest bound. This originates from the classical nature of coherent states. On the other hand, a quantum-leaning cat state seems better fitted for a quantum estimator. As one would expect, the thermal state provides no benefits, increasing the variance while decreasing the post-selection range. We attribute this result to its lack of quantum coherence in the Fock basis. Here, neither estimators (SC nor PB) approach the CR bound within an order of magnitude.   

Quantum coherence has recently been shown to be pivotal in phase estimation~\cite{Munoz_2022,Ahnefeld_2026}. Here, we follow the idea that changes in coherence of the mechanical oscillator right after post-selection will signal regions for enhanced variance. To show this, we introduce a measure of coherence susceptibility as its first derivative with respect to the phase, $|dC/d\varphi|$. $C$ is the $l_1$-norm of the off-diagonal elements~\cite{Baumgratz_2014}. In Fig.~\ref{fig_variance} (right axis) we observe that this susceptibility is a good witness for low variance regions. 

\subsection{Dissipative evolution}

In the previous subsection, we analyzed the unitary (lossless) evolution of the system governed by the unitary operator (\ref{Uop}). Nevertheless, realistic physical systems inevitably interact with their environment, and therefore an ideal lossless description is insufficient for capturing experimentally relevant behaviour. To obtain predictions consistent with realistic implementations, dissipative channels and decoherence mechanisms must be incorporated into the model.
The dynamics of the hybrid quantum system can be described within the framework of open quantum systems by means of a Markovian master equation in Lindblad form, given by
\begin{eqnarray}\label{ME1}
\frac{d\hat{\rho}}{dt}
=-i[\hat{H},\hat{\rho}]+\Gamma_s(1+\bar{n}_{s})\mathcal{L}_{\hat{\sigma}^{-}}[\hat{\rho}]+\Gamma_s \bar{n}_{s}\mathcal{L}_{\hat{\sigma}^{+}}[\hat{\rho}]+\frac{\gamma_\varphi}{2}\mathcal{L}_{\hat{\sigma}_{z}}[\hat{\rho}],
\end{eqnarray}
where $\hat{H}$  is the system Hamiltonian defined in Eq.~(\ref{Hsys}). Here, $\Gamma_s$ and $\gamma_\varphi$ are the dissipation and dephasing rates, respectively, associated with the coupling between the spin degree of freedom and a thermal reservoir characterized by a mean occupation number $\bar{n}_{s}$. We neglect MO losses under the assumption that these are weaker than those of the spin. The superoperator $\mathcal{L}_{\hat{O}}[\hat{\rho}] = \hat{O}\hat{\rho}\hat{O}^\dagger - \frac{1}{2}\{\hat{O}^\dagger\hat{O},\hat{\rho}\}$ represents the standard Lindblad dissipator, describing incoherent relaxation and thermal excitation processes.
This description assumes weak system-environment coupling and short environmental correlation times, allowing the dynamics to be treated within the Born-Markov approximation and ensuring completely positive and trace-preserving evolution.

 \begin{figure*}[h!]
 	\centering
    \includegraphics[scale=0.75]{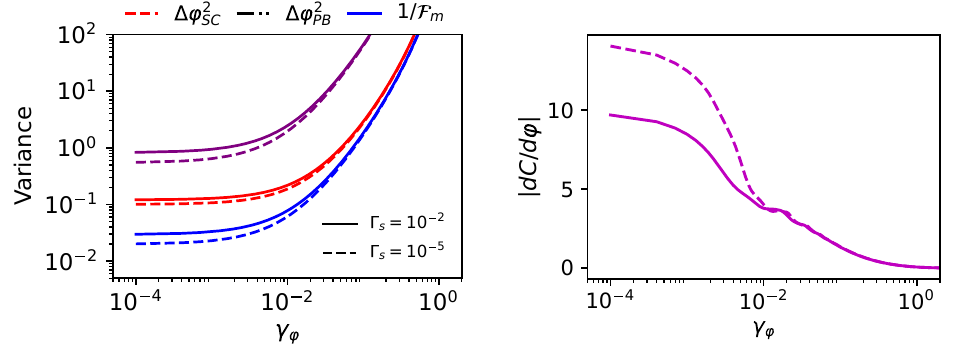}
 	\caption{Dephasing rate impairs the Variance, without a significant impact from the dissipation rate (left panel). Consistenly with our previous observations, $|dC/d\varphi|$ shows a monotonic decreasing (right panel). We simulate an initial coherent state and set $\theta_p=1.65$. $\gamma_\phi$ and $\Gamma_s$ are scaled by the oscillator frequency. Other parameters are the same as in Fig. \ref{fig_variance}.}
 	\label{fig_loss}
 \end{figure*}

To assess the impact of the environment on the estimators, we consider an initial coherent state (similar results were obtained for cat and thermal states, not shown here). As shown in Fig.~\ref{fig_loss} (left), the variances remain nearly constant in the weak-dephasing regime and then rapidly increase with $\gamma_\varphi$, indicating a progressive degradation of the estimation accuracy. Reducing the spin relaxation rate from $\Gamma_s=10^{-2}$ to $\Gamma_s=10^{-5}$ (dashed curves) leads to only a modest improvement, showing that spin dephasing is the dominant limiting factor. This behaviour is consistent with Fig.~\ref{fig_loss} (right), where the coherence susceptibility $|dC/d\varphi|$ decreases as $\gamma_\varphi$ increases.

\subsection{Role of the amplitude of initial coherent state}
\begin{figure}[ht]
\centering
\includegraphics[scale=0.24]{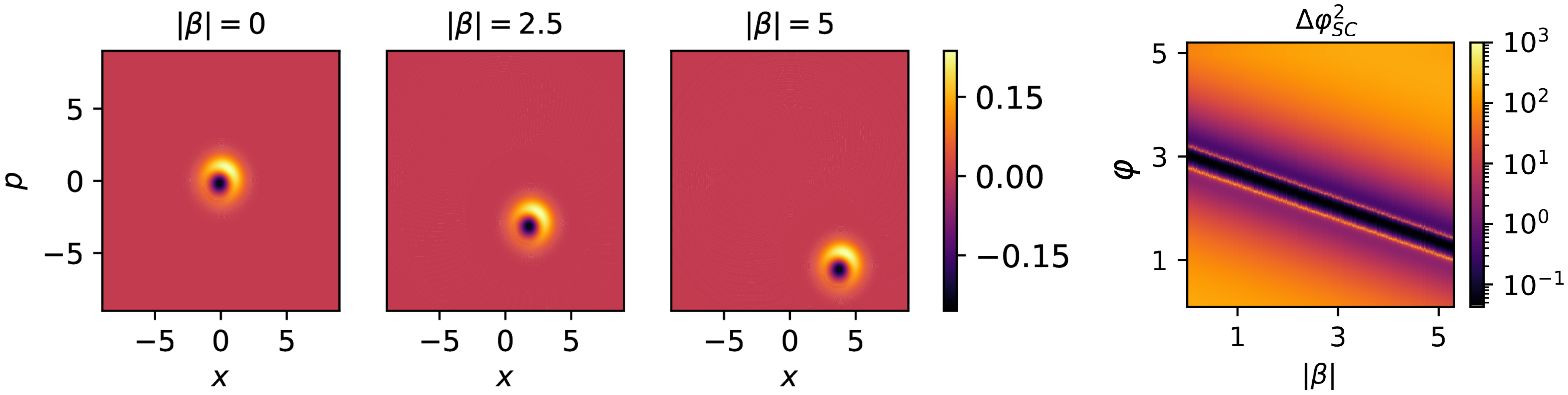}
\caption{Wigner function of the MO state is displaced for different values of the initial amplitude $|\beta|$. We set the post-selection angle at $\theta_p=1.65$, yielding a success probability of $1.5\%$ (left). $\Delta \varphi_{SC}^{2}$ as a function of the phase $\varphi$ and the coherent-state amplitude $|\beta|$ (right). Other parameters are the same as in Fig.~\ref{fig_variance}.}
\label{fig_fidelity}
\end{figure}

The amplitude $|\beta|$ of the mechanical oscillator (MO) impacts the post-selection range by displacing the MO in phase space. For illustration, let's focus on an initial coherent state and consider the SC estimator. In this case, the post-selected state of the MO approaches a displaced one-phonon Fock state, $\hat{D}(\beta e^{-i\omega_m \tau})|1\rangle=\exp{\left[\beta e^{-i\omega_m \tau}\hat{b}^{\dagger}-\beta^{*} e^{i\omega_m \tau}\hat{b}\right]}|1\rangle$, ($\hat{D}(\cdots)$ denotes the standard displacement operator) with fidelity $0.92$. The fidelity remains unchanged for a wide range of $|\beta|$. This behaviour is further clarified in Fig.~\ref{fig_fidelity} (left), which displays the Wigner functions evaluated at amplitudes $0,2.5,$ and $5$. As $|\beta|$ increases, the whole distribution is moved in phase space, preserving the negative part associated with the displaced one-phonon Fock.

Figure~\ref{fig_fidelity} (right) shows the SC variance as a function of the phase $\varphi$ and the initial coherent-state amplitude $|\beta|$. Note that steering $|\beta|$ enables control over the phase range that is intended to read.

\subsection{Enhancement by multiple spins}

Additional spins are important resources, since each spin becomes a new sensor leading to the collection of more information for the same sensing time. Hence, we now focus on transferring the phase information from several spins via post-selection. We consider a new Hamiltonian as $\hat{H} =  \omega_m \hat{b}^\dag \hat{b} + \sum_{n=1}^N(\gamma B_z\hat{\sigma}_{zn} + \lambda_0 \hat{\sigma}_{zn} (\hat{b}^\dag  +  \hat{b}))$, where $N$ is the number of spins. Without loss of generality, we consider the system initially prepared in a product state where the spins are in a superposition and the MO is prepared in a coherent state $\vert\beta\rangle$, expressed as $\vert\Psi_i\rangle=\frac{1}{\sqrt{2}^{N}}\otimes_{k=1}^N\left(\up+\down\right)_k\otimes\vert\beta\rangle$. After the evolution for a time $\tau$, we post-select each spin individually onto the state $\vert\psi_p\rangle=\left(\cos{(\theta_p/2)}\up+\sin{(\theta_p/2)}\down\right)$.

\begin{figure}[ht]
\centering
\includegraphics[scale=0.75]{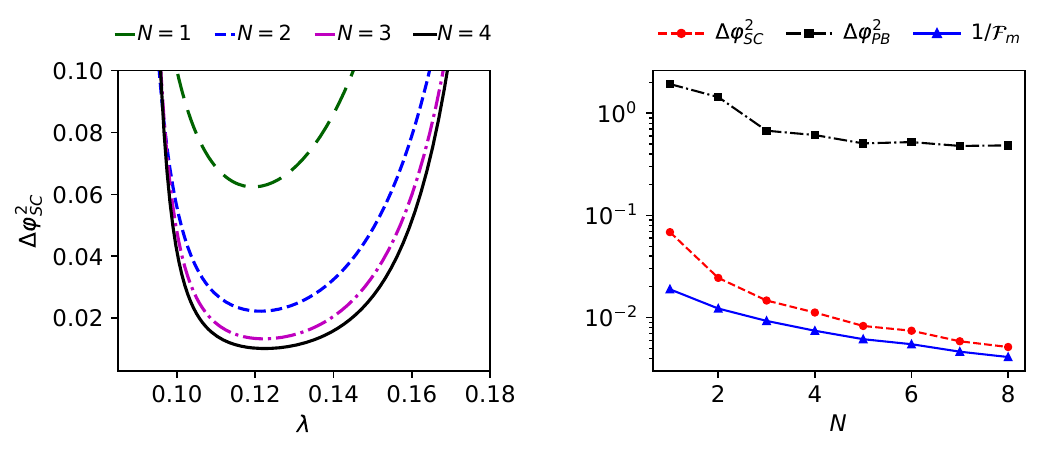}
\caption{Variance $\Delta \varphi_{\mathrm{SC}}^{2}$ shows an optimal region as a function of the coupling strength $\lambda$.  Different numbers of spins ($N$) coupled to the MO share the same optimal value $\lambda\simeq0.12$ (left). Variance's estimators decreases as a function of the number of spins. We set $\lambda=0.13$ (right). Other parameters are: $\beta=1$, $\omega_m \tau=1$, $\theta_p=1.65$, $\gamma/\omega_m=1$ G$^{-1}$ and $B_z = 2.7$ G.}
\label{fig_coup}
\end{figure}

In Fig. \ref{fig_coup} (left) we show the variance $\Delta \varphi_{\mathrm{SC}}^{2}$ as a function of the coupling strength $\lambda$ for different numbers of spins coupled to the MO. The scaled spin-mechanics coupling $\lambda$ exhibits an optimal region where it can transfer the phase information from the spin to the mechanics. It is worth noting that the optimal coupling ($\lambda\simeq0.12$) does not depend on the number of spins ($N$). Moreover, the variance rapidly saturates with $N$, as shown in Fig.~\ref{fig_coup} (right), suggesting that improvements can be attained over a few spins. Interestingly, this behaviour with the coupling strength was also observed on the mean number of phonons while cooling the mechanical oscillator~\cite{Montenegro3}.

We would like to remark that collective measurements, for instance in Bell basis, show no improvement over independent measurements.

\section{Conclusions}\label{Sec_conclusions}

Our results clarify how post-selection redistributes phase information from a spin system to a mechanical oscillator (MO). Rather than focusing on whether post-selection improves magnetometry accuracy, we have identified key factors that affect phase information transfer.  We found that increasing the number of spins ($N$) leaves invariant the optimal spin-mechanics coupling. On the other hand, the variance seems to saturate with $N$, suggesting that a small number of spins is enough for an enhancement in variance. We also found that a semiclassical variance estimator approaches the Cram\'er-Rao bound for a coherent state (classical) of the MO, while the Pegg-Barnett quantum estimator outperforms for a Schr\"{o}dinger-cat state (quantum-leaning). This provides operating regions for the estimators and the initial state of the MO. Furthermore, we introduced a coherence susceptibility that signals low variance regions based on the variation of quantum coherence. It also reveals that the dephasing mechanism hurts the variance greater than the dissipation mechanism. In addition, this work also provides a theoretical framework for exploring probabilistic phase estimation protocols in high-dimensional hybrid systems.

\section{Acknowledgments}
We are grateful to V. Montenegro for early discussion and helpful comments about this project.

\section{Authors contribution}
R.C. conceived the original idea, performed preliminar simulations and carried out the analytical calculations. H.M. and V.E. developed the python code that was finally used in all simulations. H.M. conducted the parameter analysis, the exploration of the operating regimes, and generated the figures, with support from R.C. and V.E. The manuscript was written by R.C. with input from all authors. All authors discussed the results and contributed to the final version of the manuscript.

\section*{Appendix}
\appendix

\section{Saturation of Quantum Fisher Information}\label{appendix_fisher}

Following Refs.~\cite{Combes,Alves}, the total QFI available after post-selection can be decomposed into two distinct contributions. Specifically,
\begin{equation} \label{QFIps}
 		\mathcal{F}_{ps}(\varphi)= F_p(\varphi) + p(\varphi)\mathcal{F}_m(\varphi),
\end{equation} 
where the first term, $F_p(\varphi)$, corresponds to the classical Fisher information associated with the probability $p(\varphi)$ of obtaining the successful post-selection outcome. It is given by
\begin{equation}
F_p(\varphi)=p(\varphi)^{-1}\left(\partial_\varphi p(\varphi)\right)^2.
\end{equation}
The second term in Eq. (\ref{QFIps}) accounts for the information contained in the post-selected meter state, weighted by the probability of the favorable outcome. Here, $\mathcal{F}_m(\varphi)$ denotes the QFI associated with the normalized meter state resulting from the successful post-selection, given in Eq.~(\ref{mechanical_state}). Since the optimal measurement performed on the mechanical oscillator is independent of the classical statistics of the post-selection process, the quantity $\mathcal{F}_m(\varphi)$ can be directly evaluated by substituting the state (\ref{mechanical_state}) into the QFI expression given in Eq.~(\ref{QFI}). This decomposition clearly separates the information encoded in the post-selection probability from that contained in the quantum state of the meter, allowing a transparent assessment of the uncertainty of the phase once it was transferred to the MO.

\begin{figure}[ht]
	\centering
	\includegraphics[scale=0.72]{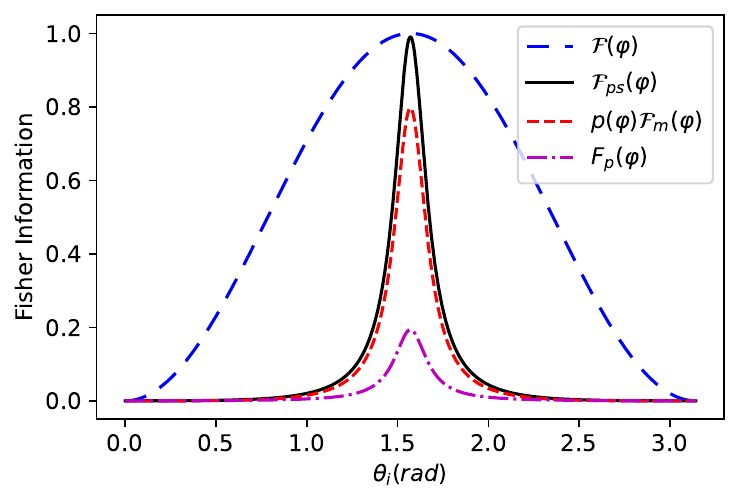}
	\caption{Quantum Fisher information (QFI) with post-selection $\mathcal{F}_{ps}(\varphi)$ saturates QFI without post-selection $\mathcal{F}(\varphi)$, in a small region of the spin preparation angle $\theta_i$. $\mathcal{F}_{ps}(\varphi)$ involves contributions from the information obtained from the meter conditioned on successful post-selection $p(\varphi)\mathcal{F}_m$, while $F_p$ denotes the classical Fisher information associated with the post-selection probability distribution $p(\varphi)$. Parameters used are $\lambda=0.1$, $\beta=1$, $\varphi=2.9\,\mathrm{rad}$, $\omega_m t=1$, and $\theta_p=\theta_i$.}
	\label{fig_ps}
\end{figure}

We found before that the maximum retrievable information from the phase amounts to $\mathcal{F}(\varphi)=\sin^2(\theta_i)$. In Fig.~\ref{fig_ps} we show that the post-selection strategy $\mathcal{F}_{ps}(\varphi)$ saturates this QFI, evidencing that post-selection is not necessarily suboptimal, although it reduces the range for initial preparation. It also illustrates how the phase's information is distributed between the post-selection probability and the MO state.

%


\section{Variation of quadrature $\hat{X}$}\label{appendix_calculations}

\begin{figure}[h]
	\centering
	\includegraphics[scale=0.74]{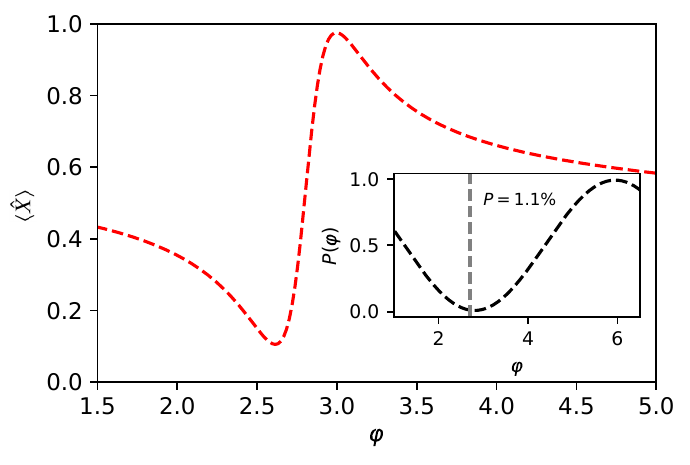}
	\caption{Strong variation of $\langle \hat{X}\rangle$, leading to high sensitivity of the MO, is impaired by the low probability of having a successful post-selection.}
	\label{fig_x}
\end{figure}

In Fig.~\ref{fig_x} we plot $\langle \hat{X}\rangle$ as a function of $\varphi$. One can observe regions where $\langle \hat{X}\rangle$ exhibits strong variations for small changes in the phase. At first glance, this behavior could be interpreted as an enhancement of the sensitivity of the mechanical oscillator (MO), which plays the role of the meter. However, this apparent advantage is accompanied by a significant reduction in the success probability $p(\varphi)$, as shown in the inset of Fig.~\ref{fig_x}. This trade-off imposes a balance between the gain in sensitivity and the probability of obtaining a successful post-selection outcome. 
For instance, for the parameters $\beta=1$, $\lambda=0.1$, $\omega_m\tau=1$, and $\varphi=2.9\,\mathrm{rad}$, we obtain a success probability of $p(\varphi)=1.1\%$ while the estimated variance is $\Delta\varphi_{SC}^2=0.1$. Importantly, this value of $\Delta\varphi_{SC}^2$ does not account for the small success probability, nor for the large fraction of experimental data discarded due to unsuccessful post-selection events. When these effects are properly included, the apparent improvement in sensitivity is significantly reduced.

\bibliographystyle{MSP}
\bibliography{biblio}

\end{document}